\documentclass[
    ,final            % use final for the camera ready runs
  ]
  {aipproc}

\layoutstyle{6x9}

\begin{document}

\title{The Light and Heavy Scalars in Unitarized Coupled Channel and Lagrangian Approaches}

\author{Frieder Kleefeld}{
  address={Centro de F\'{\i}sica das Interac\c{c}\~{o}es Fundamentais (CFIF), Instituto Superior T\'{e}cnico,\\
Edif\'{\i}cio Ci\^{e}ncia, Piso 3, Av. Rovisco Pais, P-1049-001 LISBOA, Portugal\\
e-mail: kleefeld@cfif.ist.utl.pt}
}

\begin{abstract} 
Using ideas underlying the flavour-blind ``Nijmegen Unitarised Meson Model'' (NUMM) \cite{ruppthesis1,vanBeveren:bd,VanBeveren:ea} we try to understand on the basis of a system of Schr\"odinger equations with one meson-meson and one (spinless) quark-antiquark channel coupled by a simple delta-shell transition potential the formation of (e.g.\ scalar) meson-meson scattering singularities in the complex momentum and energy plane. Surprisingly we are able to describe without direct meson-meson interaction and without any need for glueballs the whole known scalar meson spectrum. ``Light'' scalar mesons (e.g.\ $f_0(600)$, $\kappa(800)$, $f_0(980)$, $a_0(985)$, $D^\ast_0(2290)$, $\ldots$) are identified to belong to the spectrum of the transition potential, while ``heavy'' scalars (e.g.\ $f_0(1370)$, $K^\ast_0(1430)$, $f_0(1500)$, $f_0(1710)$, $a_0(1450)$, $D^\ast_0(2621)$ (?), $D^\ast_0(2825)$ (?), $D^\ast_{sJ}(2928)$ (?), $\ldots$) are related to the confinement spectrum. Due to the particular value of the charm-strange reduced quark mass level-(anti)crossing in the complex momentum plane \cite{Hernandez:xy} occurs which relates the BABAR state $D_s(2317)$ \cite{Aubert:2003fg} to the bare groundstate of the confinement spectrum, while the respective groundstate of the transition potential ends up as $D^\ast_{sJ}(2782)$ (?). We conclude with a short comment on (our) recent progress in the consistent quantum field theoretical effective description of resonances within a Lagrangian framework \cite{Kleefeld:2002au,Kleefeld:2003xy,Kleefeld:2001xd}.
\end{abstract}

\maketitle

%%%%%%%%%%%%%%%%%%%%%%%%%%%%%%%%%%%%%%%%%%%%
%% MAINMATTER
%%%%%%%%%%%%%%%%%%%%%%%%%%%%%%%%%%%%%%%%%%%%

Hadronic excitations with scalar quantum numbers are a topic of heated dispute \cite{Tuan:2003bu}. Unitarized coupled channel approaches as the one discussed here (see e.g. Ref.\ \cite{vanBeveren:2002gy} and references therein) are particularly useful to understand the non-perturbative formation and nature\footnote{``Heavy'' scalars are --- disregarding glueballs --- mainly associated to the spectrum of the confining quark-antiquark interaction used (e.g. a harmonic oscillator potential), while ``light'' (dynamically generated \cite{vanBeveren:2002mc}) scalars were identified by the author (see e.g.\ the comments in Refs.\ \cite{vanBeveren:2002gy,vanBeveren:2003uh}) to belong to the spectrum of the meson-(anti)quark transition potential (being to a good approximation of ${}^3P_0$-type).} 
 of (e.g.\ scalar) meson-meson scattering singularities in the complex momentum or energy plane, which then may enter as effective degrees of freedom with complex mass and coupling parameters effective Lagrangians describing --- according to the ``bootstrap'' idea --- meson-meson scattering already at tree level\footnote{The Lagrangian of the Quark-Level Linear Sigma Model (QLL$\sigma$M) \cite{Delbourgo:dk,vanBeveren:2002mc,Scadron:2003yg} (and references therein) has not only been shown to be an excellent canditate to achieve this task as it reproduces with a minimum of parameters at tree-level a broad spectrum of experimental facts (including the correct prediction of the mass of the now experimentally confirmed $\kappa(800)$-meson) and allowed us to gain some insight in the quark-content of scalar mesons \cite{Kleefeld:2001ds}, yet could be also ``derived'' \cite{Kleefeld:2002au} from the Lagrangian of QCD. This ``derivation'' shows that ``glueballs'' and ``quark-antiquark excitations'' seem to be synonymous.}. For a spherically symmetric situation we couple --- in the simplest case --- one Schr\"odinger equation describing a ``bound'' quark-antiquark system (confining potential $V_B(r)$) to one Schr\"odinger equation describing a meson-meson scattering continuum by a transition potential denoted by $V_T(r)$. I.e., we consider the the following coupled system of radial Schr\"odinger equations ($k := k_S := (2\,\mu_S \, (E-E^{\,(0)}_S))^{1/2}$, $k_B := (2\,\mu_B \, (E-E^{\,(0)}_B))^{1/2}$):
\begin{eqnarray} \left( d^2/dr^2 - L (L + 1)/r^2- 2\, \mu_S \; V_S (r) \; + k^2_S \, \right) \, \psi_S (r) & = & 2\, \mu_S \; V_T (r) \; \, \psi_B (r) \, , \nonumber \\
\left( d^2/dr^2 - \, \ell (\ell + 1)\,/r^2 \, - 2\, \mu_B \; V_B (r) + k^2_B \right) \, \psi_B (r) & = & 2\, \mu_B \; V_T (r) \; \, \psi_S (r) \; ,\nonumber \end{eqnarray}
with $\psi_S (0) = 0$ and  $\psi_B (0) = 0$. Even though a majority of publications is trying to find the source of ``light'' scalars in the meson-meson scattering potential $V_S (r)$, we are disregarding this interaction in what follows completely (i.e. we set $V_S (r)=0$). The conveniently normalized\footnote{Orthonormality: $\int_0^\infty dr \; \phi^\ast_{m,\ell}(r) \; \phi_{n,\ell}(r) = \delta_{mn}$; completeness: $\sum_{n,\ell} \, \phi_{n,\ell}(r) \; \phi^\ast_{n,\ell}(r^\prime) = \delta (r - r^\prime)$.} eigensolutions $\phi_{n,\ell}(r)$ of the ``bound'' system for vanishing transition potential (i.e.\ $V_T (r)$=0) correspond to the respective eigenvalues $k_{B,n,\ell} = (2\,\mu_B \, (E_{B,n,\ell}-E^{\,(0)}_B))^{1/2}$.  
After integrating the ``bound'' problem using a Green function\footnote{Green function: $G_{\ell} (r,r^\prime;E-E^{(0)}_B) =  - 2\, \mu_B \; \sum_{n,\ell} \phi_{n,\ell}(r) \; \phi^\ast_{n,\ell}(r^\prime) \; (k^2_B - k^2_{B,n,\ell} + i \, \varepsilon)^{-1}$.} $G_{\ell} (r,r^\prime;E-E^{(0)}_B)$ and reinserting it into the ``scattering'' problem we arrive at the following generalized scattering problem (with $E=(k^2+m^2_1)^{1/2}+(k^2+m^2_2)^{1/2}\,$): \\
\noindent$(d^2\!/dr^2 - L (L + 1)/r^2 + k^2) \psi_S (r) = - 2 \mu_S \sum_{\ell} V_T(r) \!\int_0^\infty \! dr^\prime  G_{\ell}\!(r,r^\prime;E-E^{(0)}_B) \, V_T (r^\prime) \, \psi_S (r^\prime)$. 
Now we will approximate astonishingly well the ${}^3P_0$ transition potential by $V_T (r) \; = \; 2\, g_T \; (2\,\mu(E)/(2\,\mu_S))^{1/2} \; \delta(r - a)$ with $2\,\mu(E) = \partial k^2/\partial E = (E^4-(m^2_1-m^2_2)^2)/(2\, E^3)$ being the relativistic meson-meson phasespace, and hence reduce the generalized scattering problem to a (radial) scattering problem at an effective $\delta$-shell described by the Schr\"odinger equation
$K^2 \; \psi_{\,L}\, (\rho) = \left( \,  - \, d^2/d\rho^2 + L (L + 1)/\rho^2 + g \;\, \delta \left(\rho - 1\right) \, \right) \, \psi_{\,L}\, (\rho)$ ($*$) with $\rho := r/a$, $K := a\,k$, $\psi_{\,L}\, (\rho) := \psi_{S , \, L}\, (r)$. 
With $\lambda_{\,\ell} \; := \; 2 g_T \; (a\; G_{\ell} (a,a;0))^{1/2}$ and $B_{n,\ell} \;  := \; (a\,\phi_{n,\ell}(a) \, \phi^\ast_{n,\ell}(a))/(a\; G_{\ell} (a,a;0))$ the dimensionless coupling $g$ displays the structure of the ``Resonance Spectrum Expansion'' (RSE) \cite{vanBeveren:2001kf} of Rupp \& van Beveren:
\begin{eqnarray} g & = & 2 \,\mu(E) \, \sum\limits_{\ell} \, \lambda^2_{\,\ell} \, \sum\limits^\infty_{n=0} \; \frac{B_{n,\ell}}{E - E_{B,n,\ell}}
 \simeq 2 \,\mu(E)  \, \sum\limits_{\ell} \lambda^2_{\,\ell} \, \, \Bigg( \sum\limits^N_{n=0} \; \frac{B_{n,\ell}}{E - E_{B,n,\ell}} - 1 \Bigg) \; . \nonumber
\end{eqnarray}
By construction there holds $\sum_{n=0}^\infty \,B_{n,\ell}\,(E^{(0)}_B - E_{B,n,\ell})^{-1} = - 1$.
The original idea of the RSE \cite{vanBeveren:2001kf} was to consider the parameters $B_{n,\ell}$, $E_{B,n,\ell}$, $\lambda_{\,\ell}$ as free parameters to fit selective meson spectra conveniently. Empirically it has become clear 
\cite{vanBeveren:2001kf,vanBeveren:2003kd}
that the parameters $\lambda_{\,\ell}$ and $B_{n,\ell}$ should be considered as ``universal'' for many different meson spectra provided the product $a \sqrt{\mu_B}$ is kept ``universal''\footnote{To keep  $a \sqrt{\mu_B}$ constant became clear from calculations performed in e.g.\ Ref.\ \cite{ruppthesis1,vanBeveren:bd} based on a transition potential $V_T(r) = g \, \omega \, \rho^{-1}_0 \; \delta(r \sqrt{\mu_B \,\omega} - \rho_0)\; \overline{V}_{int} = g\,(\mu_B \,a)^{-1} \; \; \delta(r - a)\; \overline{V}_{int}$ successfully applied to a wide range of vector meson spectra. This transition potential was a simplified version of the harmonic oscillator form of the ${}^3P_0$ transition potential inferred by G.\ Rupp \cite{ruppthesis1} and successfully applied within the NUMM \cite{ruppthesis1,vanBeveren:bd,VanBeveren:ea} in the representation $[V_T(r)]_{ij} = \tilde{g} \, \omega \;  c_{ij} \; (E/E^{(0)}_S)^{1/2} \; (r/r_0) \; \exp(\! - \, (r/r_0)^2/2\,)$ with $r_0 :=\rho_0\,(\mu_B \,\omega)^{-1/2}$. The flavour-blindness \cite{vanBeveren:2002gy} of QCD is reflected here not only by the recoupling coefficients $c_{ij}$ (or $\overline{V}_{int}$) \cite{ruppthesis1,vanBeveren:bd,VanBeveren:ea,vanBeveren:1998qg} but also by the ``universal'' values for $\rho_0$ and $\omega$ ($(\mu_B \,\omega)^{-1/2}$ has the meaning of an oscillator length). In the RSE $(E/E^{(0)}_S\, )^{1/2}$ finds the interpretation of the square root of the relativistic meson-meson phasespace, as $4\,\mu(E)/E^{(0)}_S = (E^4-(m^2_1-m^2_2)^2)/(E^3 \,(m_1+ m_2)) \stackrel{m_1=m_2}{\longrightarrow} E/E^{(0)}_S$.}. In the results displayed below we adopted the philosophy of the NUMM to describe all meson-spectra on the basis of a harmonic oscillator potential\footnote{For the harmonic oscillator potential $V_B(r)=\frac{1}{2}\,\mu_B \,\omega^2 \,r^2$ we have $E_{B,n,\ell} - E^{(0)}_B = \omega \, \left(2\, n + \ell + \frac{3}{2}\right)$ and
\[B_{n,\ell} (\bar{\rho}) = \frac{2\,\omega \; \sqrt{\pi}}{\Gamma \left(\frac{1}{2} \left(\ell + \frac{3}{2}\right) \right)\,\Gamma \left(\frac{1}{2} \left(\ell + \frac{5}{2}\right) \right)}  \;  \frac{\left(\bar{\rho}^2/2\right)^{\ell + \frac{1}{2}} \, e^{\,- \bar{\rho}^2}}{I_{\frac{1}{2} \left(\ell + \frac{1}{2}\right)} \left(\bar{\rho}^2/2\right)\,  K_{\frac{1}{2} \left(\ell + \frac{1}{2}\right)} \left(\bar{\rho}^2/2\right)} \; \frac{\Gamma\left(\ell + \frac{3}{2}\right) \; n!}{\Gamma\left(n+ \ell + \frac{3}{2}\right) } \, \left| L^{\left(\ell + \frac{1}{2}\right)}_n(\bar{\rho}^2) \right|^2\, .\]
Here we defined an ``universal'' parameter $\bar{\rho}:= a\; (\mu_B \,\omega)^{1/2}$, while $I_\nu(z)$ and $K_\nu(z)$ are modified Bessel functions, and $L^{(\alpha)}_n(z)$ are standard generalized Laguerre polynomials.} with an ``universal'' oscillator frequency $\omega=190$~MeV, and constituent quark masses \cite{Scadron:2003yg} $m_u=m_d=337$~MeV, $m_s=1.44 \, m_u$, $m_c\simeq m_D= 1865$~MeV. Hadronic resonances in meson-meson scattering are then determined for Eq.\ ($*$) by solving the respective resonance condition $\cot \delta_{L} = i$ with\footnote{S-wave ($L=0$) meson-meson scattering yields $i = - \cot K - K /(g \; \sin^2 K) \;\Leftrightarrow \; g= 2 \,i\, K/(1- \exp(2\,i\,K))$.} $\cot \delta_{L} = (n_{L} (K)/j_{L} (K)) - K / (g \; (j_{L} (K))^2)$. As in Ref.\ \cite{vanBeveren:2001kf} we will choose for the description of scalar mesons ($L=0$) a RSE with two P-wave ($\ell=1$, $N=1$) bare quark-antiquark ($\,q\,\bar{q}^{\,\prime}\;$) states in each meson-meson ($\,M \, M^{\,\prime}\,$) scattering channel. Hence, the respective RSE resonance condition to be solved in each meson-meson scattering channel is\footnote{Here we defined $\lambda:=\lambda_1$, $E^{\; q\,\bar{q}^{\,\,\prime}}_{M \, M^{\,\prime}} (K):= ((K/a_{q\,\bar{q}^{\,\,\prime}})^2+m^2_M)^{1/2}+((K/a_{q\,\bar{q}^{\,\,\prime}})^2+m^2_{M^\prime})^{1/2}$, and $\mu^{\; q\,\bar{q}^{\,\,\prime}}_{M \, M^{\,\prime}}(K) := ((K/a_{q\,\bar{q}^{\,\,\prime}})^2+m^2_M)^{1/2} \; ((K/a_{q\,\bar{q}^{\,\,\prime}})^2+m^2_{M^\prime})^{1/2}/E^{\; q\,\bar{q}^{\,\,\prime}}_{M \, M^{\,\prime}} (K)$. \\
Note that $a_{n\bar{n}} \,\sqrt{\mu_{u\bar{u}}}=a_{s\bar{s}} \, \sqrt{\mu_{s\bar{s}}}=a_{u\bar{s}} \,\sqrt{\mu_{u\bar{s}}}=a_{c\bar{s}} \,\sqrt{\mu_{c\bar{s}}}=a_{c\bar{d}} \, \sqrt{\mu_{c\bar{d}}}$ with $\mu_{q\,\bar{q}^{\,\,\prime}} := m_q \,m_{\bar{q}^{\,\,\prime}}/(m_q + m_{\bar{q}^{\,\,\prime}})$.}
\[ \frac{2 \,i\, K}{1- \exp(2\,i\,K)}
 \simeq 2 \,\mu^{\; q\,\bar{q}^{\,\,\prime}}_{M \, M^{\,\prime}}(K)  \, \lambda^2 \, \, \Bigg( \frac{B_{0,1}(\bar{\rho})}{E^{\; q\,\bar{q}^{\,\,\prime}}_{M \, M^{\,\prime}}(K) - E_{B,0,1}}+ \frac{B_{1,1}(\bar{\rho})}{E^{\; q\,\bar{q}^{\,\,\prime}}_{M \, M^{\,\prime}}(K) - E_{B,0,1}-2\,\omega} - 1 \Bigg) \; . \] 
Up to now the bare groundstates of the harmonic oscillator have to be determined empirically. Here we choose\footnote{Note that $k\simeq 766$~MeV $\simeq 4\omega$ resulting from $E_{B,1,1}$ seems to be to a good approximation ``universal''.} $E_{B,0,1}=1310$~MeV \cite{vanBeveren:2001kf} for S-wave $\pi\pi$-, $KK(I=0)$-, \mbox{$\pi K$-}, $\pi\eta_{n\bar{n}}$-scattering, $E_{B,0,1}=2440$~MeV for S-wave $D\pi$-scattering, and $E_{B,0,1}=2545$~MeV \cite{vanBeveren:2003kd} for S-wave $D K$-scattering. 
$a_{u\bar{s}}$ and $\bar{\rho}$ are then determined such that for given $\lambda$ the mesons $\kappa(800)$ (pole-position $(714-i\,228)$~MeV \cite{vanBeveren:2001kf}) and $f_0(980)$ (pole-position $980$~MeV) are reproduced simultaneously. In a good approximation $f_0(980)$ is here assumed to be purely strange \cite{Kleefeld:2001ds}. In using $m_\pi=140$~MeV and $m_K=494$~MeV we obtain the approximate result $a_{u\bar{s}}\simeq 2.55357 \;\mbox{GeV}^{-1}$ and $\bar{\rho}\simeq 1.45555$ yielding immediately $B_{0,1}(\bar{\rho}) \simeq 0.285546 \; \mbox{GeV}$, $B_{1,1}(\bar{\rho}) \simeq 0.0166127\; \mbox{GeV}$, and $\lambda \simeq 1.11572 \; \mbox{GeV}^{-1/2}$. On the basis of these parameters and $m_{\eta_{n\bar{n}}}=757.9$~MeV (for a mixing angle of $41.84^\circ$ in $n\bar{n}$-$s\bar{s}$ basis \cite{Klabucar:2001gr}) we can determine the solution of further selective RSE resonance conditions of interest in choosing e.g.\ $M,M^\prime\in\{\pi,\eta_{n\bar{n}},K,D,\ldots\}$\footnote{For S-wave $\pi\pi$-, \mbox{$KK(I=0)$-}, $\pi K$-, $\pi\eta_{n\bar{n}}$-, $D\pi$-, and $DK$-scattering we find the following pole positions in the complex energy plane: 
\underline{$\pi\pi$-scattering:} $(516 - i\;412) \; \mbox{MeV}$ ($f_0(600)$), $(1385 - i\;81) \; \mbox{MeV}$ ($f_0(1370)$, $E^{dressed}_{0,1}$), 
$(1694 - i\;4) \; \mbox{MeV}$ ($f_0(1710)$, $E^{dressed}_{1,1}$);
\underline{$KK(I=0)$-scattering:} $980 \; \mbox{MeV}$ ($f_0(980)$), $351 \; \mbox{MeV}$ (Virtual BS), $(1452 - i\;191) \; \mbox{MeV}$ ($f_0(1500)$, $E^{dressed}_{0,1}$),
$(1692 - i\;11) \; \mbox{MeV}$ ($f_0(1710)$, $E^{dressed}_{1,1}$);
\underline{$\pi K$-scattering:}  
$(721 - i\;215) \; \mbox{MeV}$ ($\kappa(800)$), $(1404 - i\; 130) \; \mbox{MeV}$ ($K_0^\ast (1430)$, $E^{dressed}_{0,1}$), $(1694 - i\; 7) \; \mbox{MeV}$ ($E^{dressed}_{1,1}$); 
\underline{$\pi \eta_{n\bar{n}}$-scattering:} $(960 - i\; 107) \; \mbox{MeV}$ ($a_0(985)$), $(1423 - i\; 161) \; \mbox{MeV}$ ($a_0 (1450)$, $E^{dressed}_{0,1}$), $(1693 - i\; 8) \; \mbox{MeV}$ ($E^{dressed}_{1,1}$);
\underline{$D\pi$-scattering:} $(2073 - i\; 70) \; \mbox{MeV}$ ($D^\ast_0(2290)$), $(2621 - i\; 163) \; \mbox{MeV}$ ($E^{dressed}_{0,1}$), $(2825 - i\; 13) \; \mbox{MeV}$ ($E^{dressed}_{1,1}$); 
\underline{$DK$-scattering:} $(2782 - i\; 166) \; \mbox{MeV}$ ($D^\ast_{sJ}(???)$), $2244\; \mbox{MeV}$ ($D_s(2317)$, $E^{dressed}_{0,1}$), $1907 \; \mbox{MeV}$ (Virtual BS, $E^{dressed}_{0,1}$), $(2928 - i\; 20) \; \mbox{MeV}$ ($E^{dressed}_{1,1}$).}. For S-wave $\pi K$-, $D\pi$-, and $DK$-scattering the results are illustrated graphically in Fig.\ 1. 
\begin{figure}
\parbox{\textwidth}{
\includegraphics[width=.48\textwidth,height=.37\textwidth]{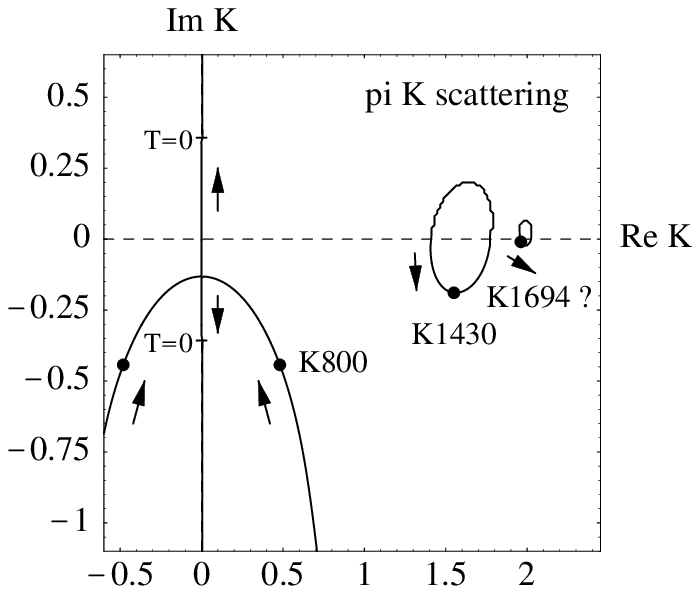}
\hfill \includegraphics[width=.48\textwidth,height=.37\textwidth]{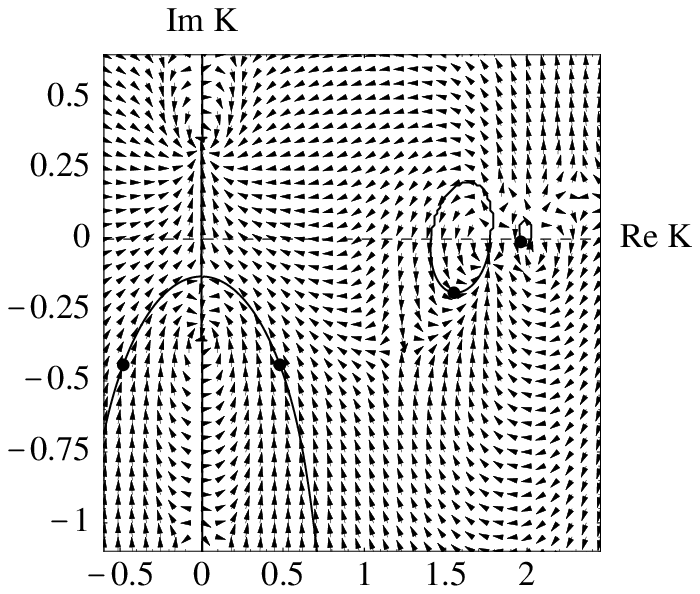}
\includegraphics[width=.48\textwidth,height=.37\textwidth]{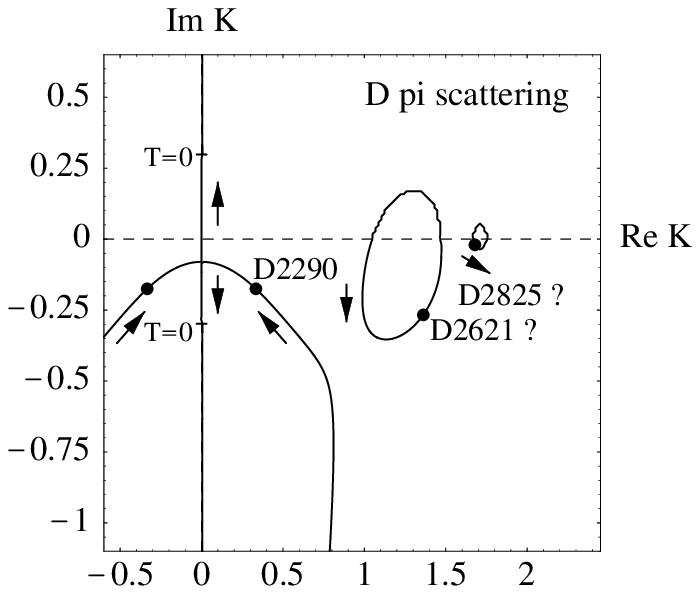}
\hfill \includegraphics[width=.48\textwidth,height=.37\textwidth]{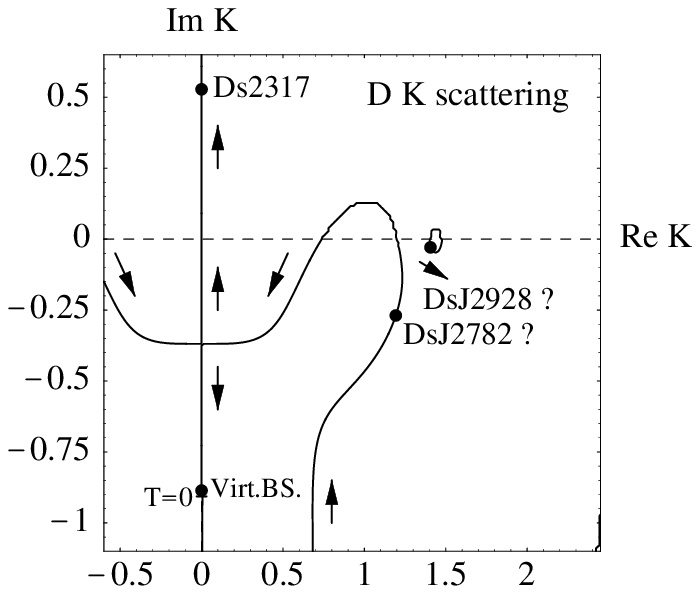}
}
\caption{Propagation of poles for increasing $\lambda^2$ in the complex $K$-plane. Solid lines: curves with Im$[\lambda^2]=0$. ``$T=0$'' indicates a zero of the amplitude due to vanishing phasespace ($\simeq$ ``Adler-zero''). Resonances at $\lambda \simeq 1.11572 \; \mbox{GeV}^{-1/2}$ are denoted by $\bullet$. S-wave scattering: \underline{$\pi K$:} $\kappa(800)$, $K^\ast_0(1430)$, $K^\ast_0(1694)$ (?); \underline{$D \pi$:} $D^\ast_0(2290)$, $D^\ast_0(2621)$ (?), $D^\ast_0(2825)$ (?); \underline{$D K$:} $D_s(2317)$, $D^\ast_{sJ}(2782)$ (?), $D^\ast_{sJ}(2928)$ (?).}
\end{figure}
In all cases there starts for $\lambda=0$ ($\Rightarrow g=0$) a $\delta$-shell pole trajectory at $K=\pi/2 - i\;\infty$. For $g\rightarrow\infty$ this pole should behave like a particle in a box and end up at $K\rightarrow\pi$. Instead, it collides for increasing $\lambda$ with the pole stemming from the bare groundstate of the confinement problem and gets deflected either to the ``left'' ($\pi K$, $D\pi$) or to the ``right'' ($DK$)\footnote{Consequently, the origin of ``light'' scalar mesons, the strong distortions of the groundstates of the ``observed'' confinement spectrum, and the absence of ``light'' non-scalar mesons for realistic transition potentials (due to the centrifugal barrier) get a nice explanation. The flavour content for $f_0(600)$, $f_0(980)$, $f_0(1370)$, $f_0(1500)$ is well consistent with Ref.\ \cite{Kleefeld:2001ds}, while observed pole-positions correspond nicely to values obtained for a realistic transition potential \cite{VanBeveren:ea}. The too large imaginary part of the $f_0(600)$ pole is an artefact of the used one-channel model. Furthermore, we observe a {\em twofold nature} of the $f_0(1710)$ being at the same time $n\bar{n}$ (relation to $f_0(600)$) and $s\bar{s}$ (relation to $f_0(980)$). The BABAR state $D_s(2317)$ \cite{Aubert:2003fg} and BELLE state $D^\ast_0(2290)$ \cite{Krokovny:2002ut} are reproduced, while the respective next higherlying states are predicted to be $D^\ast_{sJ}(2782)$ (?) and $D^\ast_0(2621)$ (?).
Note also the prediction of $K_0^\ast(1694)$ (?) and $a_0(1693\ldots1694)$ (?).}.

Shortly we address effective Lagrangian approaches: scalar mesons are characterized within a Lagrangian close to bootstrap typically by complex mass and coupling parameters to be determined by a coupled channel approach. The formalism to describe fields within a non-Hermitian Lagrangian has been provided \cite{Kleefeld:2002au,Kleefeld:2003xy,Kleefeld:2001xd}. The QCD-Lagrangian has been mapped \cite{Kleefeld:2002au} into a QLL$\sigma$M-Lagrangian replacing the gluon-quark interaction $g\;\overline{q^{\,c}_+}(x) \!\not\! A(x) \, q_-(x)$ by a {\em non-Hermitian} meson-quark interaction. The resulting scalar-meson-quark interaction $i\,\sqrt{2\,N_F/N_c}\;\,g\; \overline{q^{\,c}_+}(x) \, S(x) \, q_-(x)$ yields an {\em asymptotic free} theory which is $PT$-symmetric  \cite{Bender:2003ve} admitting a {\em real spectrum} and a {\em probability interpretation}. 
%%%%%%%%%%%%%%%%%%%%%%%%%%%%%%%%%%%%%%%%%%%%%%%%
%% BACKMATTER
%%%%%%%%%%%%%%%%%%%%%%%%%%%%%%%%%%%%%%%%%%%%%%%%
\begin{theacknowledgments}
This work has been supported by the 
{\em Funda\c{c}\~{a}o para a Ci\^{e}ncia e a Tecnologia} of the {\em Minist\'{e}rio da Ci\^{e}ncia e da Tecnologia (e do Ensinio Superior)} \/of Portugal, under Grants no.\ PRAXIS
XXI/BPD/20186/99, SFRH/BDP/9480/2002, POCTI/\-FNU/\-49555/\-2002.
\end{theacknowledgments}

%% If the bibliography is produced without BibTeX comment out the

% \bibliographystyle{aipproc}   % if natbib is available
% \bibliographystyle{aipprocl} % if natbib is missing

% \bibliography{kleefeld}
{
}
%  \typeout{******************************************}
%  \typeout{** Please run "bibtex \jobname" to optain}
%  \typeout{** the bibliography and then re-run LaTeX}
%  \typeout{** twice to fix the references!}
%  \typeout{******************************************}

\end{document}